\documentclass[]{article}

\usepackage{geometry}
\usepackage[utf8]{inputenc} 
\usepackage[T1]{fontenc}    
\usepackage{hyperref}       
\usepackage{url}            
\usepackage{booktabs}       
\usepackage{amsfonts}       
\usepackage{nicefrac}       
\usepackage{microtype}      
\usepackage{lipsum}
\usepackage{authblk}
\usepackage{subcaption}
\usepackage{graphicx}
\usepackage{amsmath}
\usepackage{amssymb}
\DeclareMathOperator*{\argmax}{arg\,max}

\title{Likelihood-based approach to discriminate mixtures of network models that vary in time}

\author[a,1]{Naomi A. Arnold}
\author[a]{Ra\'ul J. Mondrag\'on} 
\author[a]{Richard G. Clegg}

\usepackage{enumitem}
\setitemize{noitemsep,topsep=0pt,parsep=0pt,partopsep=0pt}

\affil[a]{School of Electronic Engineering and Computer Science, Queen Mary University of London, London, E1 4NS}
\affil[1]{Corresponding author: \href{mailto:n.a.arnold@qmul.ac.uk}{n.a.arnold@qmul.ac.uk}}

\begin{document}

\maketitle

\begin{abstract}
Discriminating between competing explanatory models as to which is more likely responsible for the growth of a network is a problem of fundamental importance for network science. The rules governing this growth are attributed to mechanisms such as preferential attachment and triangle closure, with a wealth of explanatory models based on these. These models are deliberately simple, commonly with the network growing according to a constant mechanism for its lifetime, to allow for analytical results. We use a likelihood-based framework on artificial data where the network model changes at a known point in time and demonstrate that we can recover the change point from analysis of the network. We then use real datasets and demonstrate how our framework can show the changing importance of network growth mechanisms over time.

\end{abstract}

\section*{Introduction}
Network growth models~\cite{barabasi1999emergence,krapivsky2000connectivity,jackson2007meeting,bianconi2001competition,vazquez2003growing}, for example, the Barab\'{a}si-Albert~\cite{barabasi1999emergence} (BA) preferential attachment model, the ranking-based model by Fortunato et al~\cite{fortunato2006scale} or the Jackson-Rogers friends of friends model \cite{jackson2007meeting}, provide an iteravive process for constructing a network by adding nodes and edges, beginning with an initial seed network. These models are widely cited as a potential explanation for the properties of real-life networks, for example the tail of the degree distribution or the clustering coefficient. Typically, network growth models provide a single mechanism that applies to the whole lifetime of the graph which has the appeal that mean-field predictions of network quantities can be derived~\cite{dorogovtsev2013evolution}. However, it has become clear that real networks can exhibit change points in their evolution~\cite{peel2015detecting,darst2016detection,torres2019non} for example the Enron email network~\cite{klimt2004enron} might be expected to change its structure when it became clear that the company was in severe legal and financial trouble. It is reasonable to expect network models to change in time given that real networks are often subjected to perturbations that we would expect to affect growth (e.g. an online social network introduces a new friend recommendation algorithm, a new field emerges in a citation network). Therefore there is a clear need for a framework that can allow network growth models that vary in time and hence a need to establish the optimal parameters for such models.

This paper provides a way to generate a large family of parameterised models that change in time and encompass the majority of existing models from the literature. We use a rigorous likelihood-based framework to compare models of growth that change in time. It has previously been shown that several network growth models can be combined to form a family of different models~\cite{clegg2016likelihood}. For example, a model could be one third BA~\cite{barabasi1999emergence} two thirds random growth~\cite{callaway2001randomly}. The best mixture to fit a given target network can be obtained using likelihood techniques~\cite{clegg2016likelihood,overgoor2019choosing}. This paper makes three contributions: (i) We create an extremely rich class of models for network growth by allowing the mixture parameters to change in time. (ii) Using artificial data generated from a known time-varying model we demonstrate a likelihood based framework that can find the correct model and its change points in time that were responsible for the observed network data. (iii) Using real data we show how this framework can give insights into the different mechanisms responsible for network growth. Hence we allow the use of existing popular models, combined into a flexible time-varying framework, either as a tool to generate new types of network or as an explanatory framework to determine the growth mechanism seen in a real data set. 

In~\cite{clegg2016likelihood} we described a likelihood based method that evaluates which hypothesised model is the best explanation for an observed network. In this paper we extend this to models that vary in time. First, we summarise the method and its extension to time-varying models. Second, we demonstrate the validity of the method by showing it is possible to accurately infer preferential attachment parameters from artificially generated networks, and correctly infer the time at which a changepoint has occurred. Finally, we study four temporal network datasets comprising a citation network~\cite{gehrke2003overview}, a StackExchange forum interaction graph~\cite{paranjape2017}, a Facebook wall posts interaction graph~\cite{viswanath2009evolution} and the Enron emails dataset~\cite{klimt2004enron}, showing the best description of each network using a time-varying model comprising three simple mechanisms: preferential attachment where an individual's chance of acquiring new links is proportional to their current number of links, triangle closure, where the chance of two nodes connecting is proportional to their number of mutual connections, and a random model where all links are equally likely to occur. On the Enron email network~\cite{klimt2004enron} we show these changes in the context of documented real-life events that would be expected to influence this network.

\section*{Framework}
In this paper we develop an extension to the framework from~\cite{clegg2016likelihood} for analysing the likelihood that a given model led to a set of observations for a dynamic graph.
The models have the common structure that at each time step they produce a probability for different potential next steps in the evolution of a graph (for example the BA model gives the probability of choosing a particular node to connect to as being proportional to its degree).
By carefully constructing a likelihood from these probabilities we can deduce which hypothesised model is the most likely explanation for a set of graph observations.
In this paper we introduce the idea that this model may vary over time and investigate this through artificial and real data.
In this section we begin by describing how the framework introduced in~\cite{clegg2016likelihood} can be extended to analyse models that vary in time.
We then introduce a new measure that shows how similar two hypothetical models for graph evolution are when applied to a given graph at a given time.
The more similar two models are the harder it will be, in principle, to tell them apart.
\subsection*{Model Structure}
We consider dynamic graphs $G(t) = \left(V(t), E(t) \right)$ where $V(t)$ is the set of nodes and $E(t)$ is the set of edges at time $t$, with number of nodes $N(t) = |V(t)|$.
For simplicity we describe undirected, unweighted simple networks and consider only networks where nodes and edges are permanent once added.
The restriction to undirected and unweighted graphs is purely for clarity of explanation and is not a fundamental constraint on the framework. Assume the graph changes at some set of times $t_1, t_2, \ldots$ and let $g_{i}$ be the observation made at time $t_i$ (that is $G(t_i) = g_i$).
We will use the word \emph{increment} to describe the change in the graph from observation $g_{i-1}$ to $g_i$ that is observed at time $t_i$.
This increment is a time and a set of nodes and links that are added at that time. Let $G_i$ be a random variable representing the graph the $i$th observation. A model is a set of rules that gives the probability $\mathbb{P}(G_i = g_i | G_{i-1}=g_{i-1})$, the probability the observed graph $g_{i-1}$ will change at time $t_i$ to the graph $g_i$.

Models that describe processes for network growth can be split into two parts: the first is the \emph{operation model} that describes the type of change being made (e.g. ``add a new node and connect it to two existing nodes" or ``add one link connecting two existing nodes"; the second, the \emph{object model}, is a set of rules for exactly which entities should be chosen.
The former explains phenomena such as densification~\cite{leskovec2005graphs} (in later stages of a network's life links are more commonly made between existing nodes rather than adding new nodes, changing the average degree) and the varying rate of node and link arrival to the network~\cite{fire2020rise}.
The operation model can be directly extracted from network data provided that the time (or order) at which each node and edge is added to the network is known.
The latter has been used to explain more structural characteristics such as power-law degree distributions, high clustering coefficients and assortative/disassortative mixing. 
As an example, the Barab\'asi-Albert model~\cite{barabasi1999emergence} specifies that, starting from an initial small seed network, at each timestep a single new node is added and connected to $m$ existing nodes (operation model), and those nodes are chosen with a probability proportional to their degree (object model).
These concepts of a graph increment and an operation and object model, simply make formal the underlying assumptions of models from the literature that define how networks grow.
\subsubsection*{Operation Model}
\begin{figure}[t]
	\centering
	\begin{subfigure}{0.45\linewidth}
		\centering
	\includegraphics[scale=0.3]{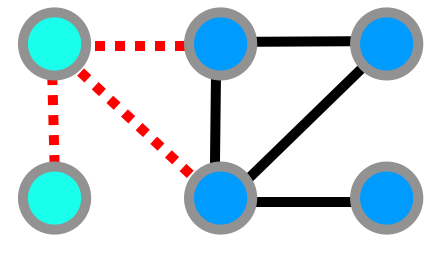}
	\caption{External Star: A new node (light blue, top left) connecting to one new and two existing nodes}
	\end{subfigure}
	\hfill
	\begin{subfigure}{0.45\linewidth}
	\centering
	\includegraphics[scale=0.3]{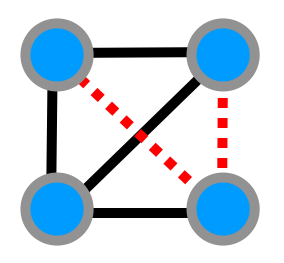}
	\caption{Internal Star: An existing node (bottom right) connecting to two other existing nodes}
	\end{subfigure}
\caption{Examples of the two types of star growth operations considered within this framework.}
\label{fig:operation}
\end{figure}

The operation model specifies the type of transformation that will happen to the graph and the time at which it will happen.
It selects the time and the number of edges and nodes and how they will be connected in the next graph increment.
In this paper we use operation models comprising growth by stars (see Figure~\ref{fig:operation} for an example), that is, a new or existing node connecting to a number of existing or new nodes.

Examples of networks which grow in this way are citation networks, in which each paper is a new node citing existing papers, or email networks, where a star is formed with the centre node the sender and outer nodes the recipients.
Another natural operation to consider is growth by cliques, characteristic of collaboration networks, which is not within the scope of this paper.

\begin{figure}[htbp]
	\centering
	\includegraphics[width=0.9\linewidth]{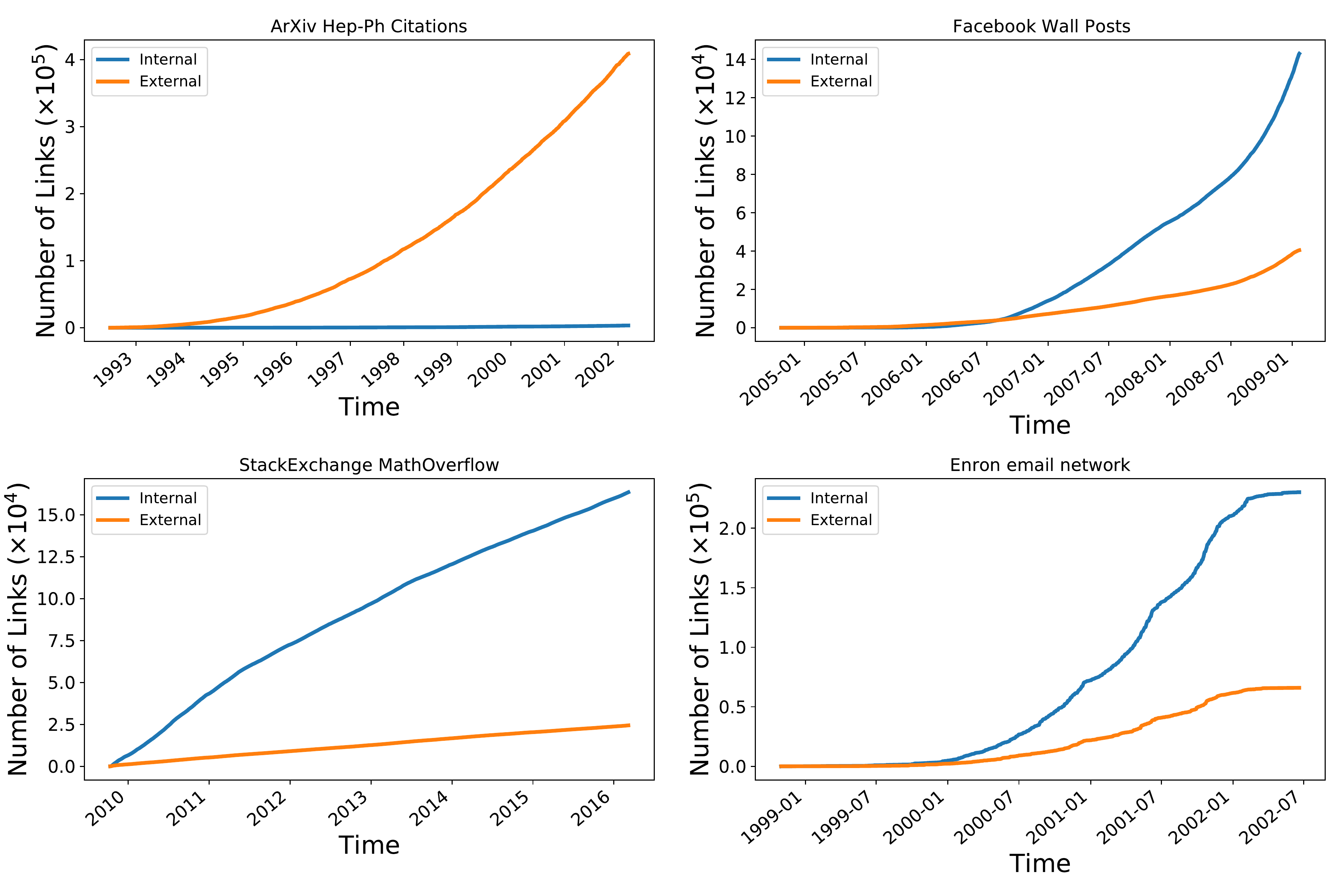}
	\caption{Number of edges in the network, split into those which join pairs of existing nodes (Internal) and those which join a new node to other new or existing nodes (External).}
	\label{fig:omreal}
\end{figure}

As an example of how the operation model affects the graph, consider Figure~\ref{fig:omreal} that separates links in a growing network into those that join new nodes to the network and those that are purely internal between nodes that already exist.
The citation network for example comprises purely external links since it always grows by a new paper citing existing papers, whereas all the others are a mixture, with most new links being internal ones.
\subsubsection*{Object model}
Given the type of transformation and number of nodes $n$ chosen by the operation model, the \emph{object model} must select which set of $n$ nodes are changed. It does this iteratively by assigning a probability $p_i(t)$ to each node $i$ in the network reflecting its chance being selected to be part of a graph increment at time $t$ with $\sum_{i=1}^N p_i(t) = 1$. This probability depends on the existing state of the graph and possibly also nodes already selected in the set. This set is selected by sampling, with or without replacement depending on whether duplicate links are allowed, from the distribution provided by the object model. This part of the model is aimed at understanding the factors governing which nodes are more likely to attract new links in a growing network. For example, imagine the operation model has selected that the next increment will be a star centred on an existing node and connecting to three other existing nodes (in an email network this represents an email sent from a user that has been previously interacted with to three other nodes that have been previously interacted with). The object model must pick a node to represent the sender and then three different nodes to represent the recipients. Some example object models (omitting the $t$ for simplicity) are:
\begin{itemize}
	\item \textbf{Random attachment} $M_{\text{rand}}$ is the simplest model and accounts for unknown behaviour not captured by other models. All nodes are assigned equal probability, so the probability of choosing node $i$ is $p_{i} \propto \text{const}$.
	\item \textbf{Degree power} $M_{\text{DP}}(\alpha)$ captures the rich-get-richer effect where nodes of a high degree are more likely to attract new links, with $p_{i} \propto k_i^{\alpha}$~\cite{krapivsky2000connectivity}. When $\alpha=1$ this corresponds to the Barab\'{a}si-Albert (henceforth BA) model\cite{barabasi1999emergence} $M_{\text{BA}}$ for generating scale-free networks. This rich-get-richer effect is amplified with larger values of $\alpha$.
	\item \textbf{Triangle Closure} $M_{\text{tri}}$ captures the tendency of triangles of connected nodes to occur. It selects edges with probabilities proportional to the number of triangles that edge would complete. The source node $j$ of an edge is chosen at random (using $M_{\text{rand}}$). The destination node $i$ is chosen with $p_i \propto |\varGamma(j) \cap \varGamma(i)|$, where $\varGamma(i)$ is the set of neighbours of node $i$. When we connect a star (as in this paper) we pick a single source $j$ using $M_{\text{rand}}$ and connect it to $n$ destinations using the probability $p_i$ defined previously.
 This is reminiscent of a friend recommendation algorithm which recommend connections between individuals with many friends in common~\cite{adamic2003friends}.
	\item \textbf{Rank-Preference model} $M_{\text{RP}}(\alpha)$ gives higher probability to nodes with higher rank. It assigns probability $p_{i} \propto R_i^{-\alpha}, \alpha>0$, where $R_i$ is the rank of node $i$ for some choice of ranking system~\cite{fortunato2006scale}. If the node labels correspond to the order in which the nodes arrive to the network, the ranking $R_i = i$ corresponds to a tendency to connect to the longer established nodes in the networks, and is an alternative mechanism to generate scale-free networks~\cite{goh2001universal}.
\end{itemize}
A network's growth may best be described by a mixture of models, and this mixture may change over time. Assume we mix together $L$ models where in the $l$th model $M_l$, the probability of selecting node $i$ at time $t$, is $p_i^{M_l}(t)$.  Therefore, we propose a mixture model with the probability of choosing node $i$ at time $t$ given by:
\begin{equation}\label{eqn:mixedmodel}
p_i(t) = \sum_{l = 1}^L \beta_{l}(t) p_i^{M_l}(t)
\end{equation}
where the sum is over the $L$ different model components $M_l$ considered, $\beta_{l}(t) \in [0,1] $ and $\sum_{l=1}^L \beta_{l}(t) = 1$.
Later in this paper we may denote a model mixture as $M(t) = \sum_{l=1}^L \beta_l(t)M_l$ as shorthand for the fuller model description given by equation~\ref{eqn:mixedmodel}.
For a fixed value of $t$ the $\beta_l(t)$ should be thought of as interpolating between the models.
For a simple concrete example, a mixture of random attachment and the standard preferential attachment model with no time-dependence on the mixture takes the form $$p_i(t) = \beta \frac{1}{N(t)} + (1 - \beta) \frac{k_i(t)}{\sum_j k_j(t)}$$ with $\beta \in [0,1]$. In this example $\beta = 0$ would give result in a pure BA model whilst $\beta = 1$ would result in a model where nodes are chosen at random. This example is investigated using a master equations approach in~\cite{ghoshal2013uncovering}.
In this paper we use values of $\beta_l(t)$ that are constant or piecewise-constant over evenly sized time intervals, and so when we are considering a model with $L$ components spanning $J$ time intervals, we may write for convenience $\beta_{lj}$ for the weight of the $l^{\text{th}}$ model component during the $j^{\text{th}}$ time interval.
\subsection*{Likelihood calculation}
Let $g_0$ be our first observation of a graph $G$, and $G_1, G_2, \dots $ be random variables representing the subsequent states of $G$, with corresponding observations $g_1, g_2, \dots $ at times $t_1, t_2, \dots$. We assume that our observations are high resolution enough that, the subgraph $g_{i} \setminus g_{i-1}$ is a small increment $\delta_{i}$ which is the set of nodes and edges added to the graph when changing from $g_{i-1}$ to $g_{i}$. (This assumption is discussed in more detail in the supplementary information.) In this sense, the random variable describing the graph at observation $n$ given a starting observation $g_0$ can be expressed as
\begin{equation}
G_n = g_0 \cup_{i=0}^n \varDelta_i.
\end{equation}
Here $\varDelta_i$ is the random variable associated with $\delta_i$, and the union $\cup$ of two graphs $G$ and $H$ should be understood as the graph whose vertex set is the union of $G$'s and $H$'s vertices and whose edge set is the union of $G$'s and $H$'s edges.

This allows us to calculate a likelihood of a model $M$ given observations $g= g_0, g_1, \dots , g_n$ of $G$ as
\begin{align}\label{eqn:likelihoodprod}
l(M| G=g) &= \prod_{i=1}^n \mathbb{P}(\varDelta_i = \delta_i | G_{i-1}=g_{i-1}, M).
\end{align}
where the probability $\mathbb{P}(\varDelta_i = \delta_i)$ is provided exactly by the object and operation model and refers specifically to the probability of selecting the nodes that are involved in that increment.
In practice, we transform equation~\ref{eqn:likelihoodprod} to use the \emph{per-choice likelihood ratio} $c_0$ from~\cite{clegg2016likelihood} given by
\begin{equation}
	c_0 = \exp \left( \frac{\log\left( l(M| G=g) \right) - \log \left( l(M_{\text{rand}}|G=g) \right)}{\sum_{i=1}^n m(i)} \right)
\end{equation}
where $m(i)$ is the number of node choices at timestep $i$.
This provides a useful reference figure of $c_0 > 1$ if the model given is more likely than the basic model $M_{\text{rand}}$ and $c_{0}<1$ if it is less likely, as well as moving the likelihood into a more human-readable range.
So if $\varDelta_i$ specifies a single link added between two existing nodes, $\mathbb{P}(\varDelta_i = \delta_i)$ for observation $\delta_i$ is the probability of selecting the observed source multiplied by the probability of the destination node from the remaining nodes.
As an explicit example, if we used the model $M_{\text{BA}}$ and our observed graph increment $\delta_i$ at time $t_i$ was the internal node with index 1 connecting to the internal node with index 2, the probability $\mathbb{P}(\varDelta_i = \delta_i)$ would be given by $\frac{k_1}{\sum_{j=1^N} k_j} \frac{k_2}{\sum_{j=2} k_j}$.
The probability $\mathbb{P}(\varDelta_i = \delta_i)$ is defined completely by the operation and object model.
The supplementary information details how this is achieved.
Hence we can calculate the likelihood of a model and rigorously compare which model from a candidate set gives the highest likelihood.
\subsection*{Model similarity}
Some pairs of models may give similar probabilities for most nodes in a graph because of properties that are correlated (e.g. the rank-preference model gives rise to a strong correlation between high rank and high degree, so gives similar node probabilities to the BA model).
To measure this, we use cosine similarity to compare the overlap in node probabilities given to a graph by a pair of different models.
The intuition is that this measure will be equal to 1 if and only if the probability distributions given to the node set by each model are identical, and closer to 0 if these distributions are very different. Let $G$ be a graph and $M_1, M_2$ be two different object models with node $i \in G$ being assigned probability $p_i^{M_1}, p_i^{M_2}$ by $M_1, M_2$ respectively.
Then we define their cosine similarity over $G$ as:
\begin{equation}
\sigma_G\left( M_1, M_2 \right) =K^{-1}\sum_{i=1}^N p_i^{M_1} p_i^{M_2}
\label{eqn:similarity}
\end{equation}
where $K= \sqrt{\left(\sum_{i=1}^N \left(p_i^{M_1}\right)^2\right) \left( \sum_{i=1}^N \left(p_i^{M_2}\right)^2 \right)}$ is a normalisation ensuring that this measure lies between 0 and 1, and is equal to 1 if and only if $p_i^{M_1} = p_i^{M_2}$ for all $i$. The numerator of this quantity is the probability that $M_1$ and $M_2$ would pick the same node of $G$ from a single draw. An important feature is that this quantity depends highly on the structure of $G$. For example, if $M_1, M_2$ are based strictly on node degree, then their similarity will be 1 if $G$ is a regular graph (all nodes of the same degree). To highlight this, consider models $M_{\text{rand}}$ and $M_{\text{BA}}$ assigning probabilities $p_i^{\text{rand}} = 1/N$ and $p_i^{\text{BA}} = k_i /\sum_{j=1}^N k_j$ respectively. Their (squared) similarity is given by:
\begin{equation}
\sigma_G(M_{\text{rand}}, M_{\text{BA}})^2 = \frac{\langle k \rangle^2}{\langle k^2 \rangle}.
\end{equation}
This is equal to 1 if and only if $G$ is regular (all nodes the same degree). On the other hand, if $G$ is close to scale-free (i.e. with $\langle k^2 \rangle$ very large compared to $\langle k \rangle$) then their similarity is close to zero.

\section*{Results}
\subsection*{Estimating preferential attachment model parameters}
To provide confidence in the likelihood framework, our first set of experiments relates to our ability to recover correct model parameters in artificial data experiments.
We estimate the preferential attachment exponent $\alpha$ in networks generated using the degree power object model $M_{\text{DP}}(\alpha)$, where each node $i$ is chosen with probability proportional to $k_i^{\alpha}$.
This is a challenging estimation problem because for $\alpha>1$ a single node eventually will attract connections from every new node that joins the network, making the difference in observed behaviour between a pair of networks grown using different but high values of alpha very small (for a more in-depth discussion on this see~\cite{falkenberg2020identifying,krapivsky2000connectivity}.)
Other estimation methods are Newman's~\cite{newman2001clustering} non-parametric method which underestimates the $\alpha$ exponent~\cite{overgoor2019choosing}.
We generate networks of 1000 nodes, with operation model comprising, at each iteration, one new node attaching to $m$ existing nodes.
We then find maximum likelihood estimates $\hat{\alpha} = \argmax_{\alpha} l(M_{\text{DP}}(\alpha)|G)$, by performing a grid search through $\alpha = -0.1, -0.09, \dots, 2.1 $.
In the case $m=1$ the likelihood can be written $l(M_{\text{DP}}(\alpha)|G) = \prod_{i=1}^t {k_{c_i}^{\alpha}}/{\sum_{j=1}^{N(i)} k_j^{\alpha}}$, where $N(i)$ is the number of nodes at time $i$ and $c_i$ is the node chosen at timestep $i$.
Instead, empirical results suggest that our method gives an estimator which does not exhibit bias and has small variance even for large powers of $\alpha$.
\begin{figure}[t]
	\centering
	\includegraphics[width=0.8\linewidth]{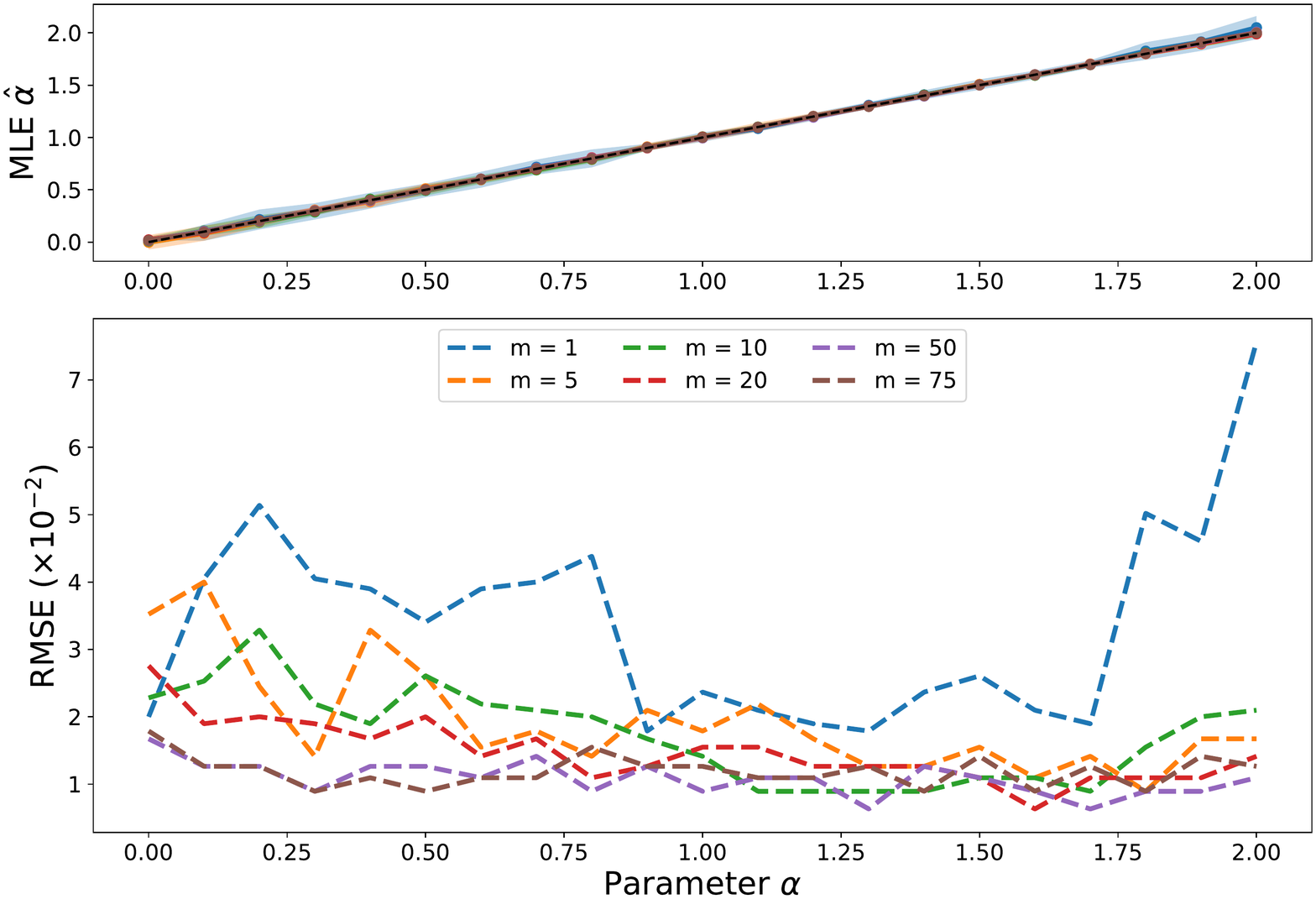}
	\caption{Estimation of non-linear preferential attachment exponent $\alpha$ with the shaded areas representing 95\% confidence intervals over ten experiments (upper) and root mean squared error in parameter estimates for $\alpha$ (lower).
	The $y=x$ line in the upper plot is included to aid comparison between our estimated value and the true values.
    For each value of $\alpha$ and $m$ the synthetic network was generated using the degree power object model $M_{\text{DP}}(\alpha)$ with each new node connecting to $m$ existing nodes; the lines and confidence intervals for each of these heavily overlap in the upper plot.}
	\label{fig:DPnochange}
\end{figure}
Figure~\ref{fig:DPnochange} shows maximum likelihood estimators of parameter $\alpha$ from artificially generated networks where a single node is added at each timestep and connects to $m$ existing nodes.
Reassuringly, increasing $m$, the number of links that arrive at each timestep in the network, does not degrade the quality of the parameter estimates.
\subsection*{Distinguishing similar network generation mechanisms}
\begin{figure}[htb]
	\centering
	\includegraphics[width=0.9\linewidth]{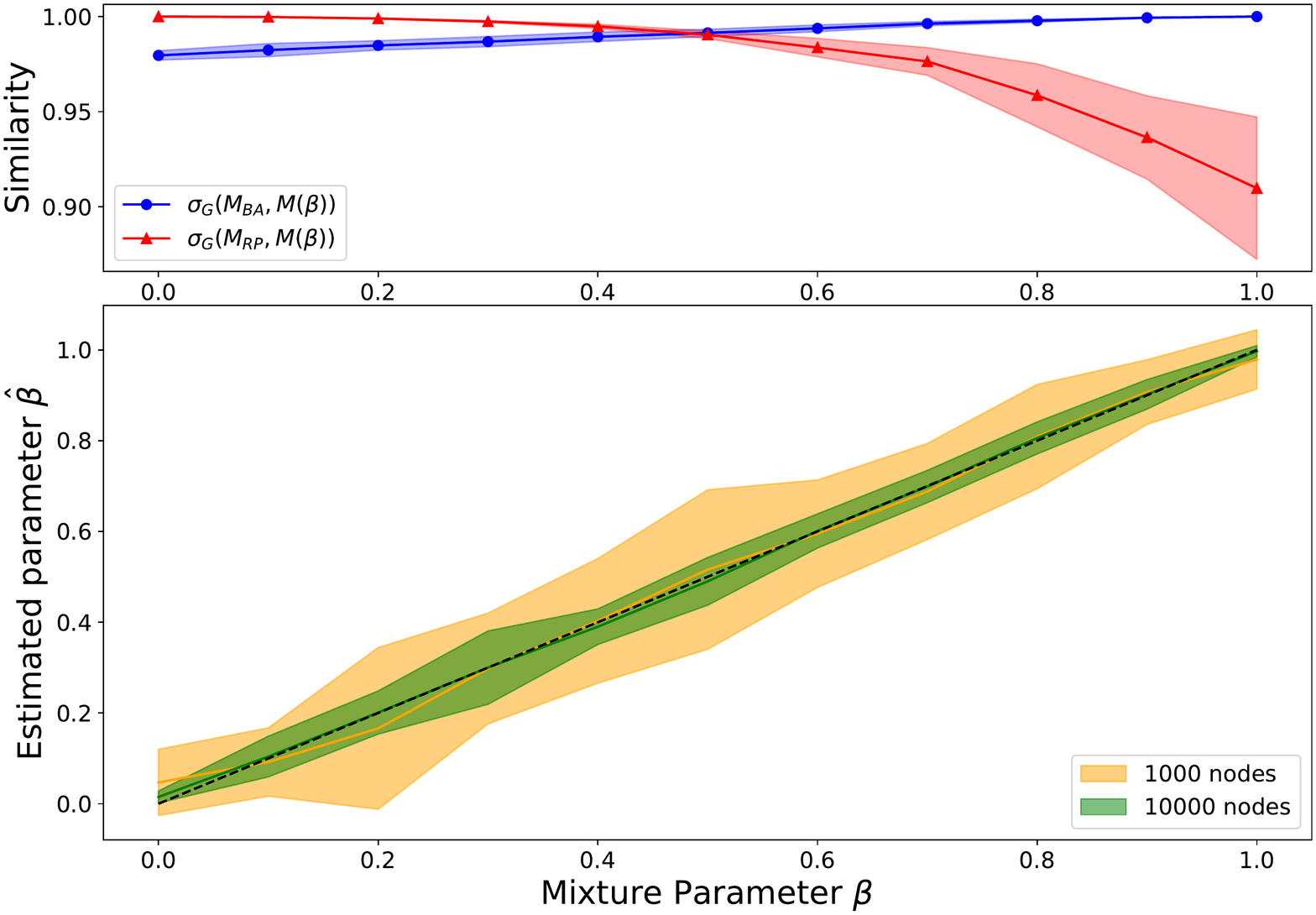}
	\caption{The upper plot shows the average model similarity values $\sigma_G\left( M_{\text{BA}}, M(\beta)\right)$ and $\sigma_G\left( M_{\text{RP}}, M(\beta)\right)$ where $G$ is the network of 1000 nodes generated from model $M(\beta)$. The lower plot shows the performance of the maximum likelihood estimator $\hat{\beta}$ in recovering the true mixture parameter $\beta$ of the model combining the BA and rank-preference mechanisms. In both plots, the shaded region represents a 95\% confidence interval over 10 repetitions, and each synthetic network was generated by, starting from a clique of 5 nodes, adding a single node at each iteration and connecting to $m=3$ existing nodes.}
	\label{fig:RPBA}
\end{figure}

As well as correctly estimating model parameters, we investigate whether our method can correctly distinguish different network generation mechanisms.
We consider a difficult task of distinguishing two models which generate networks that have very similar degree distributions and summary statistics: the BA object model which generates scale-free networks via preferential attachment to node degree and the (static) rank-preference model which achieves the same end but via preferential attachment to the oldest nodes.
To test whether we can distinguish between these mechanisms, we combine them in an object model we refer to as
\begin{equation}\label{eqn:BARP}
M(\beta) = (1-\beta) M_{\text{RP}}(0.5) + \beta M_{\text{BA}}
\end{equation}
so that $\beta=0$ gives a model that is entirely rank-preference and $\beta=1$ entirely BA.
The rank-preference model parameter $\alpha = 0.5$ is chosen to yield the same degree distribution power law exponent $\gamma=3$ as the BA model~\cite{fortunato2006scale}.
We generate networks of 1000 and 10000 nodes, with operation model comprising, at each iteration, one new node attaching to 3 existing nodes, chosen according to the object model in Equation~\ref{eqn:BARP}.
For each of the 1000 node networks, we calculated the similarity between the model $M(\beta)$ that generated it and each of the BA and the rank-preference models, finding between 90 and 100\% overlaps for each of the model pairs over the whole range (Figure~\ref{fig:RPBA} top.)
We then calculate maximum likelihood estimators $\hat{\beta}$ by performing a (parallel) search through the space $\beta = 0, 0.01, 0.02, \dots , 1$.
The mean of these maximum likelihood estimators (over 10 realisations for each parameter) and SD error areas is displayed in Figure~\ref{fig:RPBA}, showing high accuracy in detangling these two very similar mechanisms.
We find intuitively that the error is smaller for the larger network, since the likelihood is calculated from ten times more datapoints in this case. The error is smaller at the extremes $\beta = 0$ and $\beta = 1$; this is because only values $\beta \in [0,1]$ are possible, so $\hat{\beta}$ cannot overshoot at either end.
\subsection*{Recovering change-points in artificial data}
Next we demonstrate that in artificial data we can recover the known point in time where one model changes to another model.
For the following experiments, we consider an object model assigning probabilities to nodes as
\begin{equation}\label{eqn:DPchange}
p_i(t) \propto \begin{cases}
k_i^{\alpha} & t \leq T \\
k_i^{\beta} & t > T \\
\end{cases}
\end{equation}
for a single changepoint time $T$, i.e. the degree power model  $M_{\text{DP}}(\alpha)$ where the exponent changes from $\alpha$ to $\beta$ at time $T$.
We test our ability to estimate $T$ from networks generated with known $T$, assuming also that $\alpha$ and $\beta$, the degree power exponents before and after the changepoint respectively, are known.
We generate artificial networks with a single changepoint at time $T$, and find an estimator $\hat{T}$ by maximising the likelihood $ \hat{T} = \argmax_T l(G | T)$.
A first observation is that more accurate estimates are obtained in the larger network.
A subtle point is that this is not because of having a higher number of observations in the larger network; indeed, the changepoints in the larger network span the same time range (1,000) as the smaller network, but because in the larger network we are observing nodes being drawn from a more stabilised distribution later in the network's lifetime.
The second is that the changepoint resulting from the smaller parameter change (1.0 to 0.9) draws noisier estimates $\hat{T}$ than for larger parameter changes (1.2 to 1.0).

\begin{figure}[htbp]
	\centering
	\begin{subfigure}{0.45\linewidth}
		\includegraphics[width=0.95\linewidth]{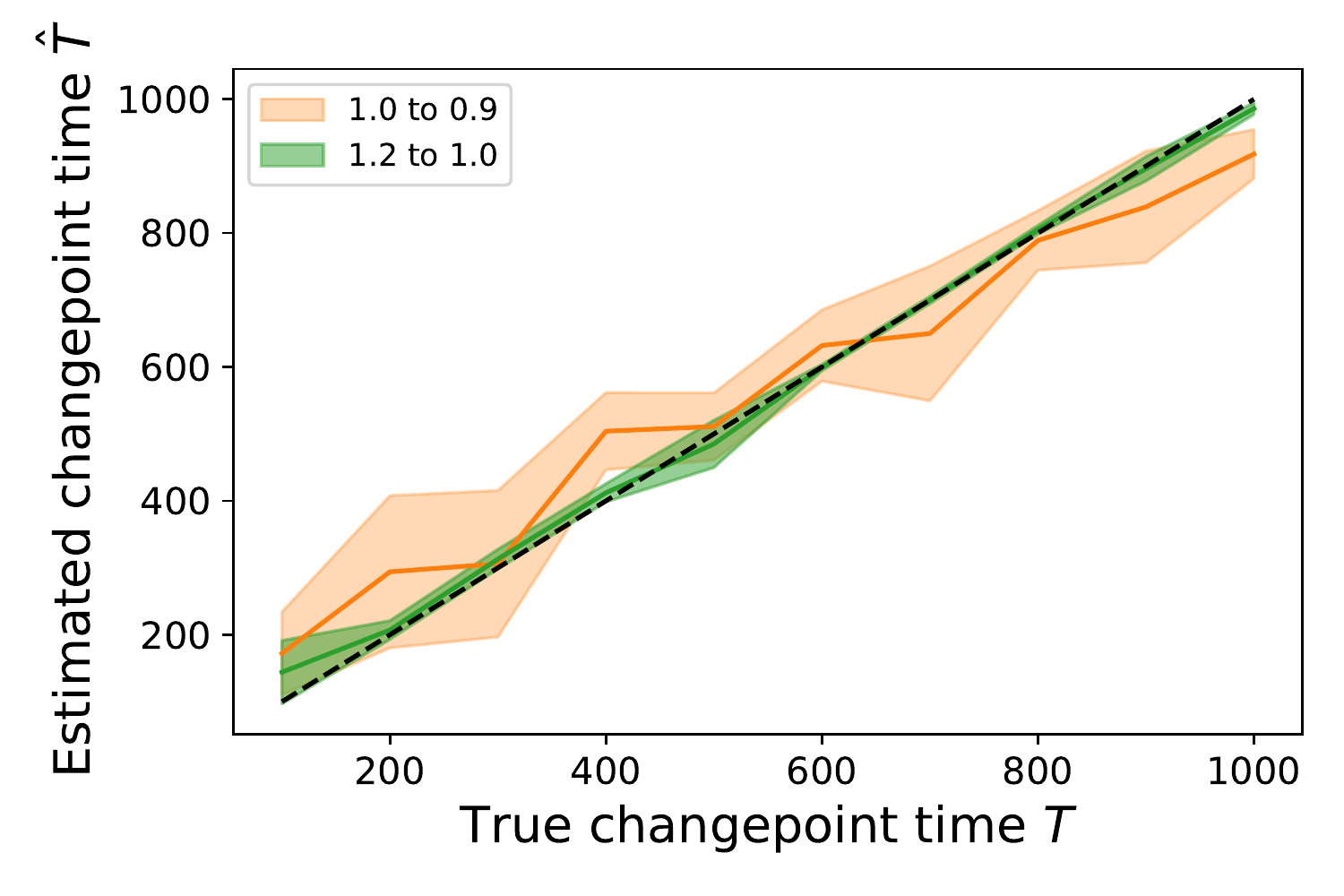}
		\caption{1,000 nodes}
		\label{fig:DP1000}
	\end{subfigure}
	\hfill
	\begin{subfigure}{0.45\linewidth}
		\includegraphics[width=0.95\linewidth]{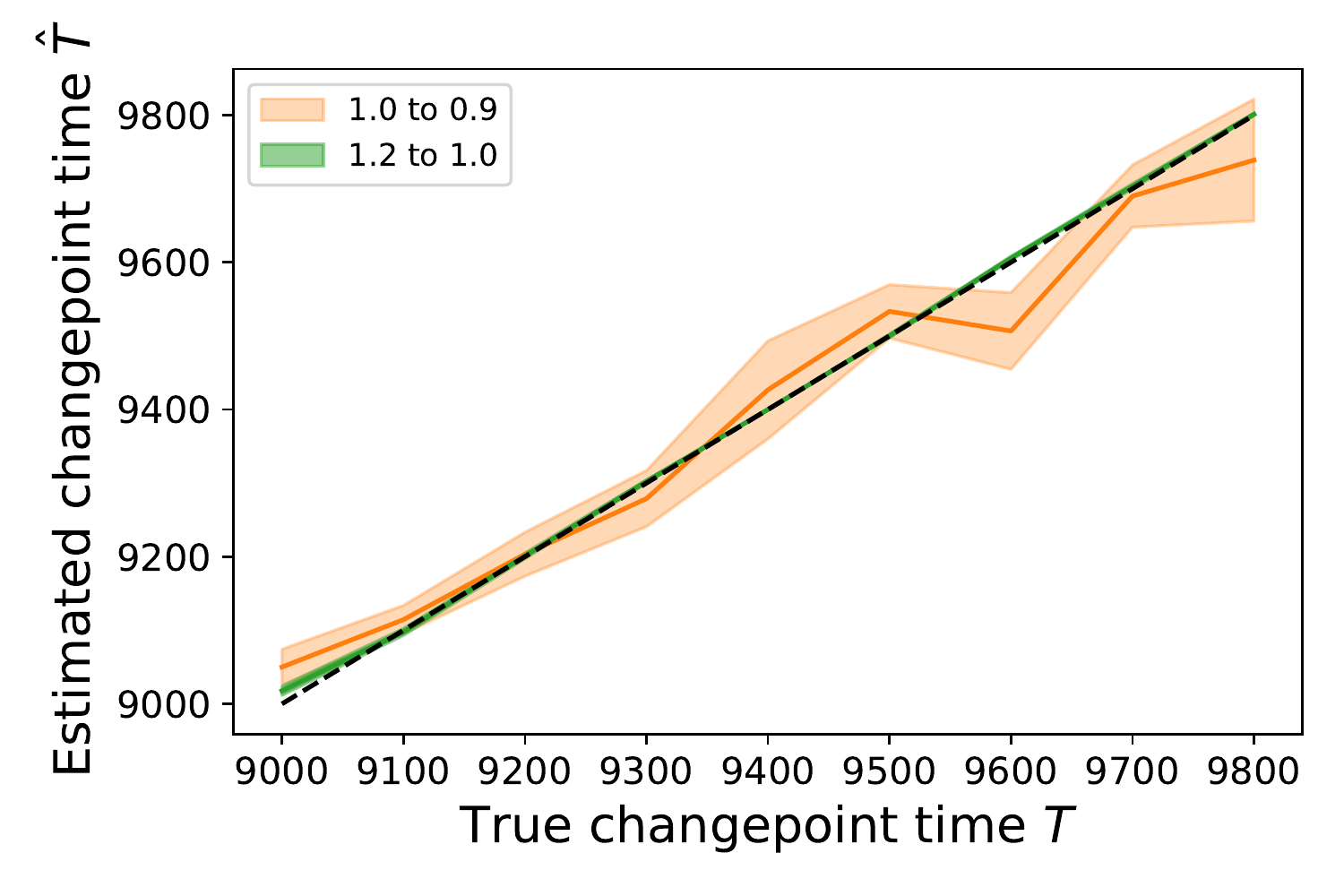}
		\caption{10,000 nodes}
		\label{fig:DP10000}
	\end{subfigure}
	\caption{Maximum likelihood estimators for time at which degree power exponent changes value, with mean and standard deviation error areas calculated over 10 experiments. In both Figures \protect\subref{fig:DP1000} and \protect\subref{fig:DP10000}, a network is generated with each new node joining connecting to 3 existing nodes, node $i$ chosen with probability $p_i$ as in Equation~\ref{eqn:DPchange}. The shaded region represents a 95\% confidence interval over the 10 experiments.}
	\label{fig:dpchanges}
\end{figure}
\begin{figure}[htbp]
	\centering
	\begin{subfigure}[t]{0.45\linewidth}
		\includegraphics[width=0.9\linewidth]{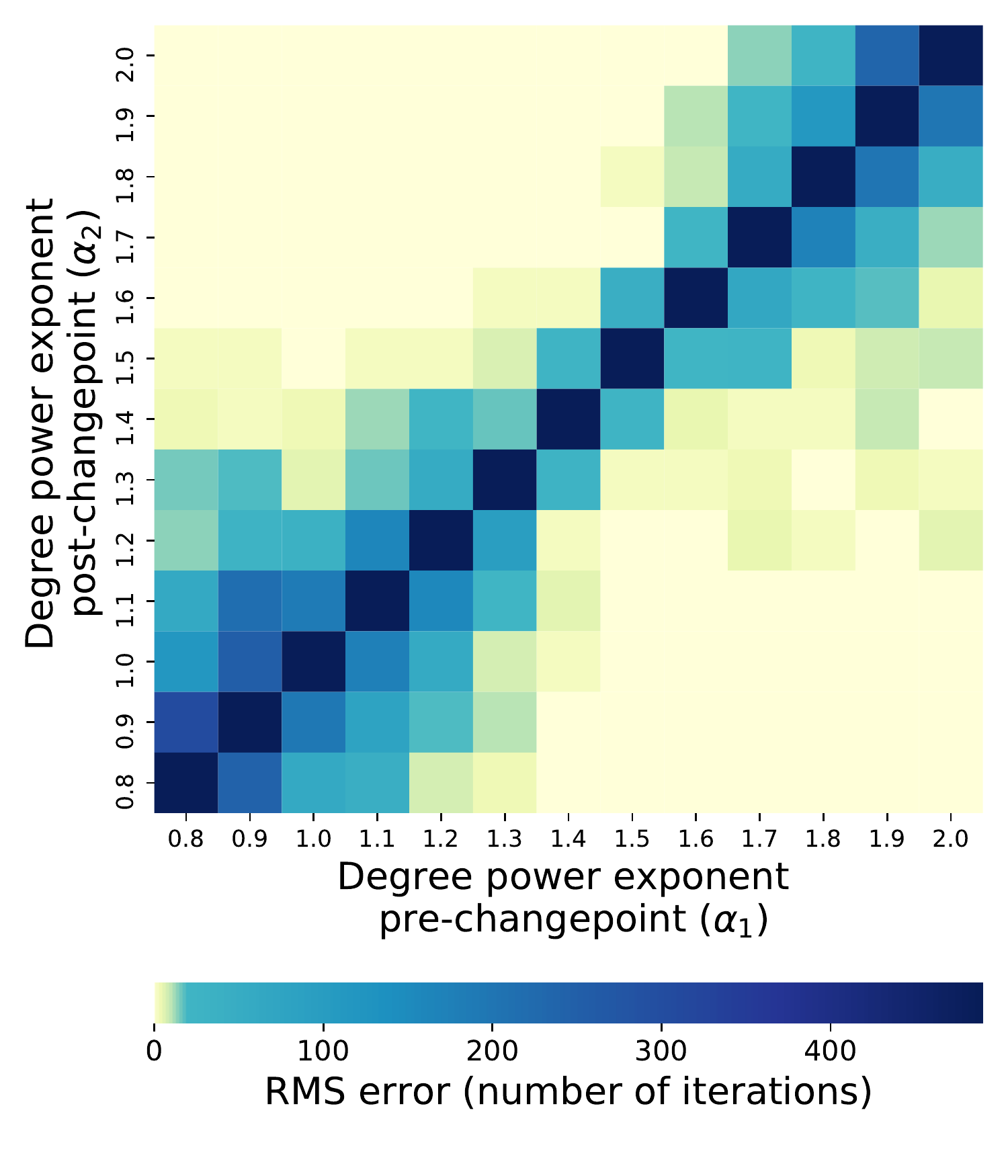}
		\caption{Heatmap showing root mean squared error (RMSE) in estimation of changepoint for different parameters pre- and post-changepoint. Importantly, the parametrisations on the diagonal correspond to `no change' and therefore have identical RMSE, equal to half the size of the time interval searched.} \label{fig:DPheatplot}
	\end{subfigure}
	\hfill
	\begin{subfigure}[t]{0.45\linewidth}
		\includegraphics[width=0.9\linewidth]{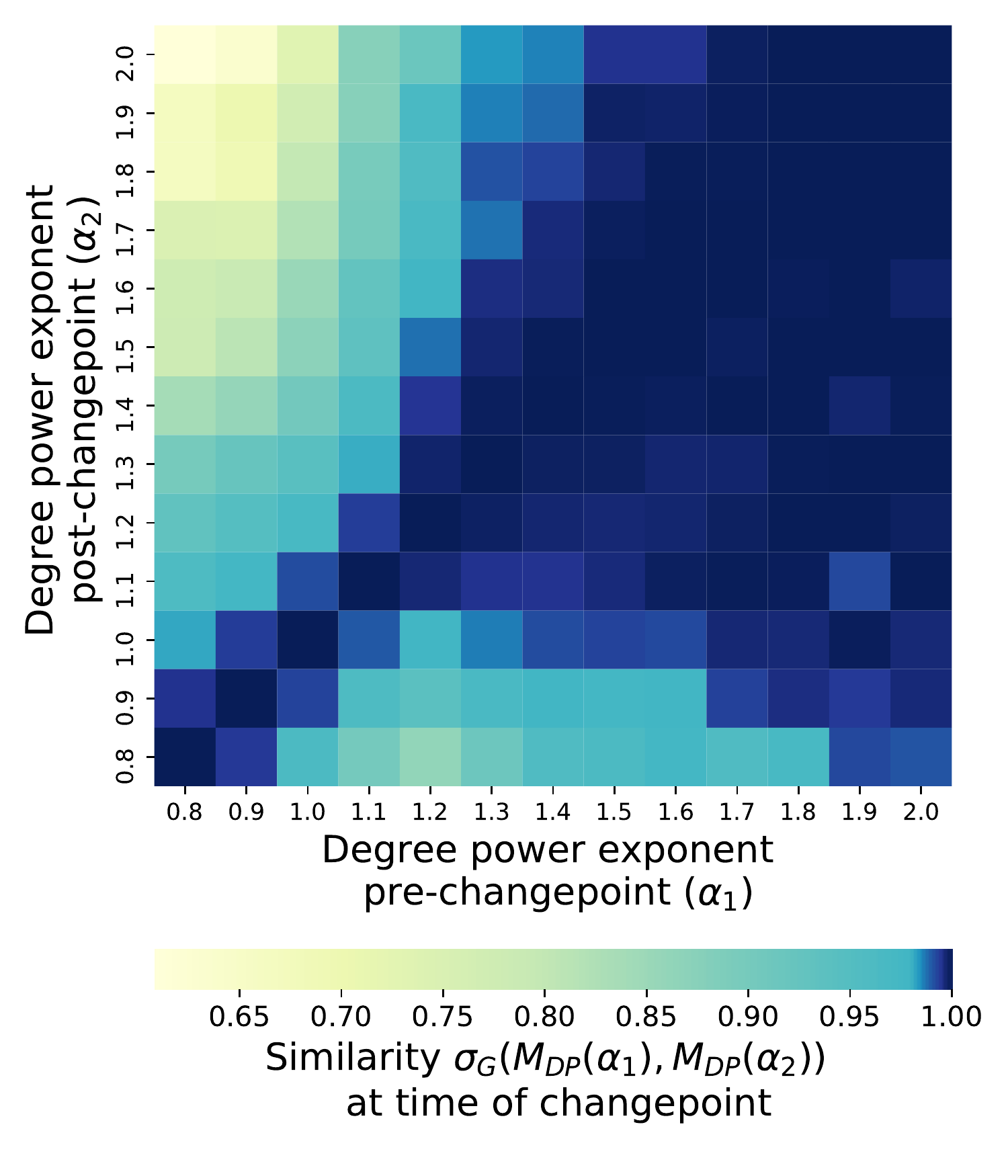}
		\caption{Heatmap showing similarity value between degree power models at the time of the changepoint with different power parameters as defined in \eqref{eqn:similarity}.} \label{fig:DPSimilarity}
	\end{subfigure}
\caption{Experiments on a 10,000 node network generated using object model described in equation~\ref{eqn:DPchange}. The operation model used is to at each iteration, add a single new node and connect to 3 existing nodes (i.e. an external star).}
\label{fig:heatmaps}
\end{figure}
We explore this latter observation in more detail by fixing our initial changepoint time $T=5,000$ and calculating the root mean squared error (RMSE) in estimates $\hat{T}$ arising from different pre- and post-changepoint parameters $\alpha_1$ and $\alpha_2$ (Figure~\ref{fig:DPheatplot}).
As with Figure~\ref{fig:dpchanges}, we note that smaller parameter changes are more difficult to detect, corresponding to the band of higher error around the diagonal $\alpha_1 = \alpha_2$.
We also notice the band increasing in width toward the extremes of the diagonal.
In the lower extreme case ($\alpha_1$ small), this is likely because the degree distribution pre-changepoint is more homogeneous and thus many nodes have a similar likelihood of being chosen.
In the higher extreme ($\alpha_1$ large) we suspect that this is due to an amplification of the rich-get-richer effect: after a certain point in time, the highest degree is so extreme that changing the power in the model slightly will not alter its near-certain probability of the corresponding node being selected.
Shown also is a heatmap (Figure~\ref{fig:DPSimilarity}) showing the model similarity calculated for pairs of models $M_{\text{DP}}(\alpha_1)$ and $M_{\text{DP}}(\alpha_2)$, over the networks generated from the model in Equation~\ref{eqn:DPchange} at the time of their changepoint.
We find asymmetry in the $\alpha_1 - \alpha_2$ phase plane, with $\sigma_G(M_{\text{DP}}(\alpha_1), M_{\text{DP}}(\alpha_2))$ being large for large $\alpha_1$ values regardless of the value of $\alpha_2$.

\subsection*{Reproduction of network statistics from synthetic networks with changepoints}
The previous test on finding the correct time of a changepoint assumed that we knew that just one changepoint was to be found and knew the form of the model before and after the changepoint.
While the results are reassuring given the discussed difficulty of preferential attachment estimation, those assumptions are unlikely to hold for a real data setting.
We test now if we can find the right number of changepoints to use and the correct models between these changepoints.
As with the previous experiment, starting from a seed network of 5 nodes we generate a synthetic target network of 10,000 nodes with a single changepoint at $T=5000$ (i.e. when the network reaches 5,000 nodes).
Specifically, we use an object model $M(t)$ of

\begin{equation}\label{eqn:BARandchange}
M(t) = \begin{cases}
				   \beta M_{\text{BA}} + (1 - \beta) M_{\text{rand}} & 0 \leq t \leq 5000 \\
				   (1 - \beta) M_{\text{BA}} + \beta M_{\text{rand}}  & t > 5000 \\
\end{cases}
\end{equation}
with $\beta=0.3$, i.e. a model that switches from being mostly random and part BA for $t \leq 5000$ and mostly random and part BA for $t>5000$, and operation model of each new node connecting to 3 existing nodes.

\begin{figure}[htb]
	\centering
	\includegraphics[width=0.9\linewidth]{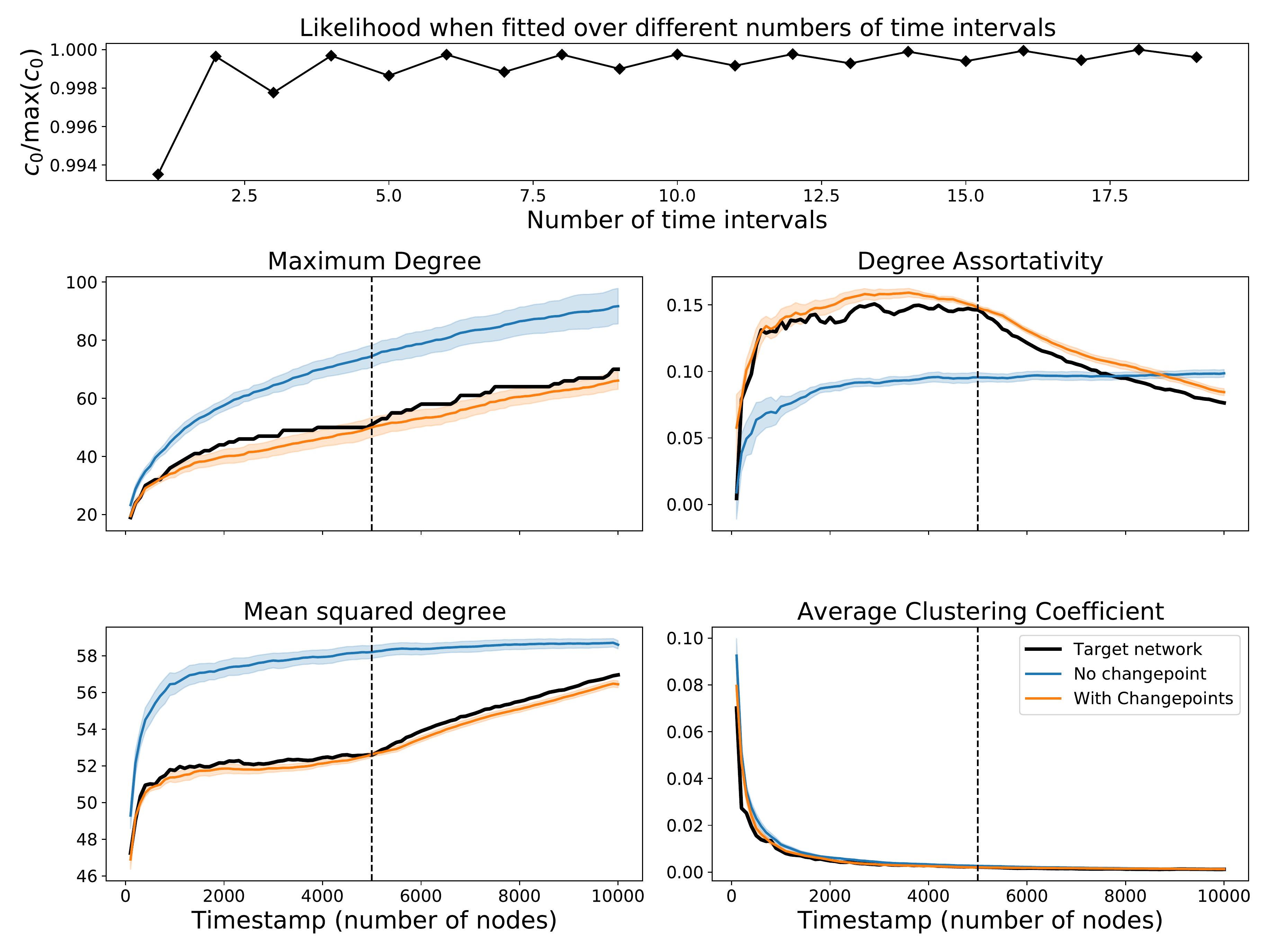}
	\caption{Investigation of a synthetic network (parameters described in text) with a single changepoint at timestamp t = 5000, signified by the dashed line in
the lower plots. The upper plot shows how the normalised model likelihood varies as the estimated number of time intervals increases (the correct number is two). The lower plot shows how well an artificial model reproduces network statistics with no changepoints or with a single changepoint. The shaded area represents a 95\% confidence interval over 10 repetitions.}
	\label{fig:artificialcomparisons}
\end{figure}

We fit a model of the form
\begin{equation}
	M(t) = M(j(t)) = \beta_{\text{rand},j} M_{\text{rand}} + \beta_{\text{BA},j} M_{\text{BA}}
\end{equation}
where $j(t)$ is the number of the interval containing time $t$, over $J$ evenly spaced time intervals $j = 1, \dots , J$, and first establish which value of $J$ works best.
The upper plot in Figure~\ref{fig:artificialcomparisons} shows how the $c_0$ value changes for numbers $J$ of time intervals.
As the synthetic network has just one changepoint exactly at the halfway point in the network's growth, the $c_0$ values alternate as $J$ moves between even and odd numbers, increasing when $J$ is incremented to an even number of intervals and decreasing when it is odd.
We then take measurements on networks generated from best fitting models for $J=1$ (no changepoints) and $J=2$ (one changepoint) compared to the target network. We consider the maximum degree $k_{\max}$, the mean squared degree $\langle k^2 \rangle$, the degree assortativity~\cite{newman2002assortative}, the average clustering coefficient and number of singleton (degree 1) nodes, all shown in Figure ~\ref{fig:artificialcomparisons} apart from the singleton nodes which were zero for all cases (since each new node has degree at least 3). For each value of $J$, we generate 10 networks and an average and confidence interval are provided.
For $J=1$ our method estimated the model to be $0.5M_{\text{BA}} + 0.5M_{\text{rand}}$ throughout; trying $J=2$ yielded an estimate of $0.29M_{\text{BA}} + 0.71M_{\text{rand}}$ for the first half and $0.69M_{\text{BA}} + 0.31M_{\text{rand}}$ for the second half.
In effect the $J=1$ model time-averages the best mixture from that of the correct number of intervals, so the measurements for these behave accordingly, with most measurements for the $J=1$ time interval model matching the target network toward the end but mismatching in the middle. 
\subsection*{Fitting time-varying models to real data}
\begin{figure}[htbp]
	\centering
	\includegraphics[width=0.9\linewidth]{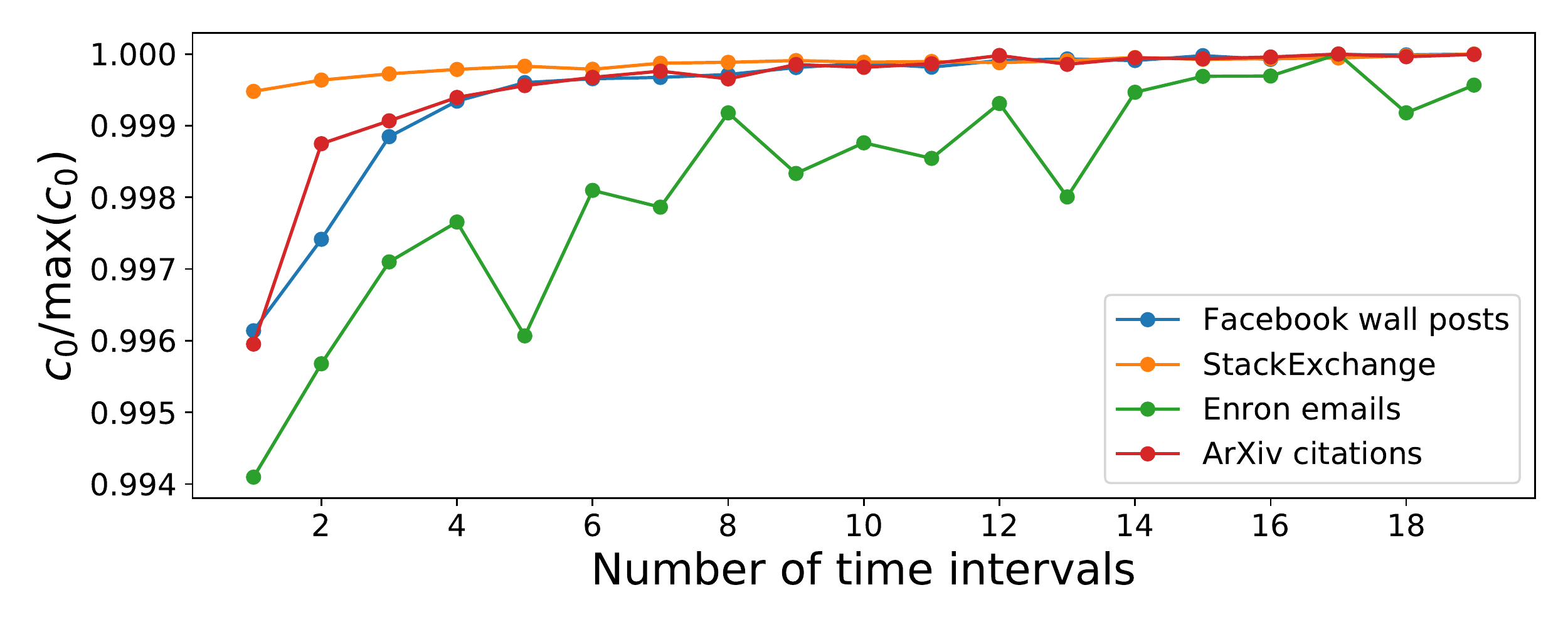}
	\caption{Likelihood ($c_0$ measure) of maximum likelihood object model of the form in equation~\ref{eqn:realdatafits}, plotted against the number of time intervals used (left) and normalised by the maximum value it takes.}
	\label{fig:c0vscps}
\end{figure}

\begin{figure}[htbp]
	\centering
	\includegraphics[width=0.9\linewidth]{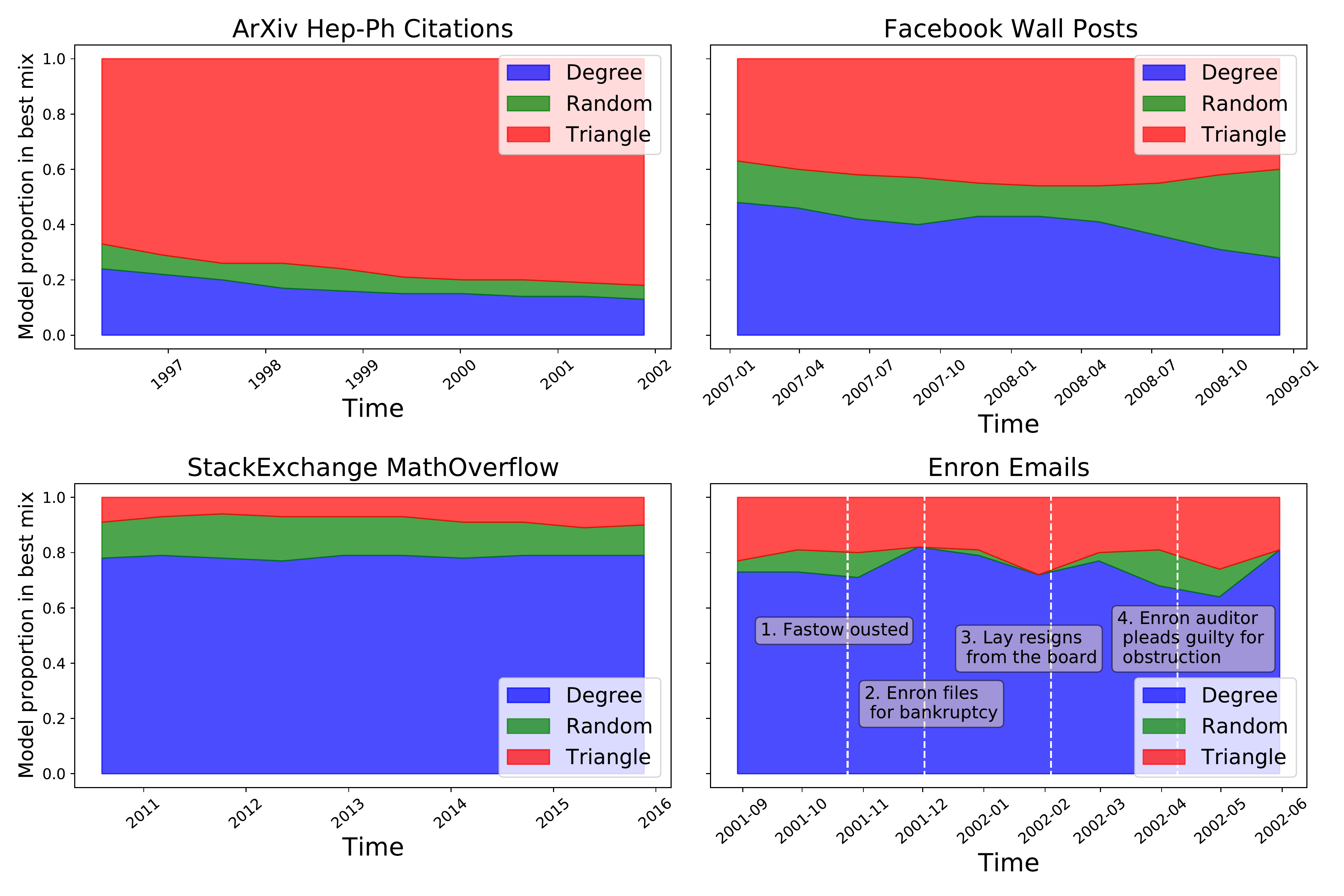}
	\caption{Best fitting (maximum likelihood) mixture of the BA, Triangle Closure and Random model over 10 time intervals.
	The Enron data is annotated with some key events in the company's collapse.}
	\label{fig:enrontime}
\end{figure}

\begin{figure*}[htbp]
	\centering
	\includegraphics[width=0.9\linewidth]{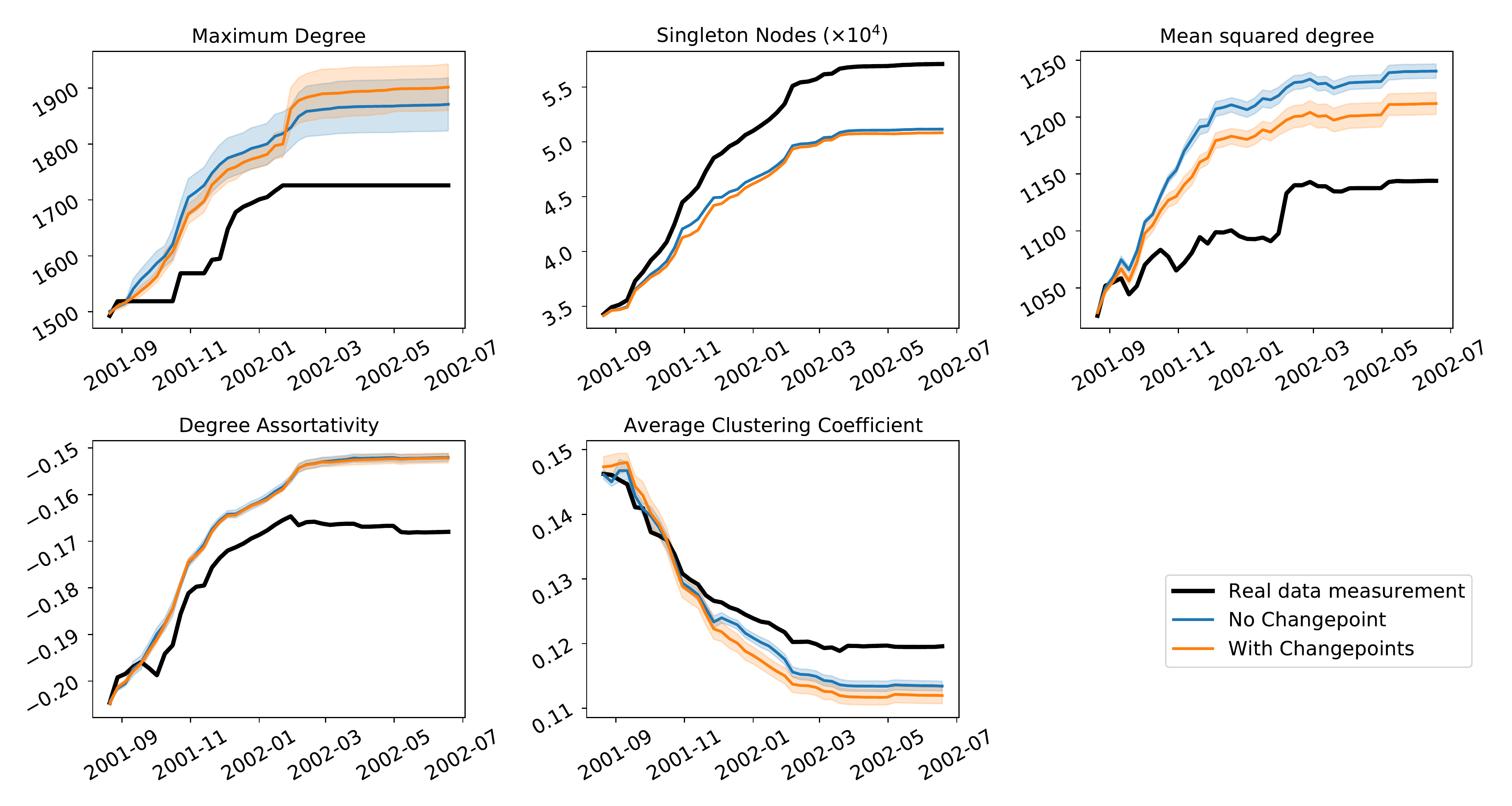}
	\caption{Enron Emails: Comparison on various network statistics between best fitting mixture model without change and its best fitting counterpart fitted over 10 equally sized intervals for the Enron emails dataset. }
	\label{fig:ENcomparisons}
\end{figure*}

\begin{figure*}[htbp]
	\centering
	\includegraphics[width=0.9\linewidth]{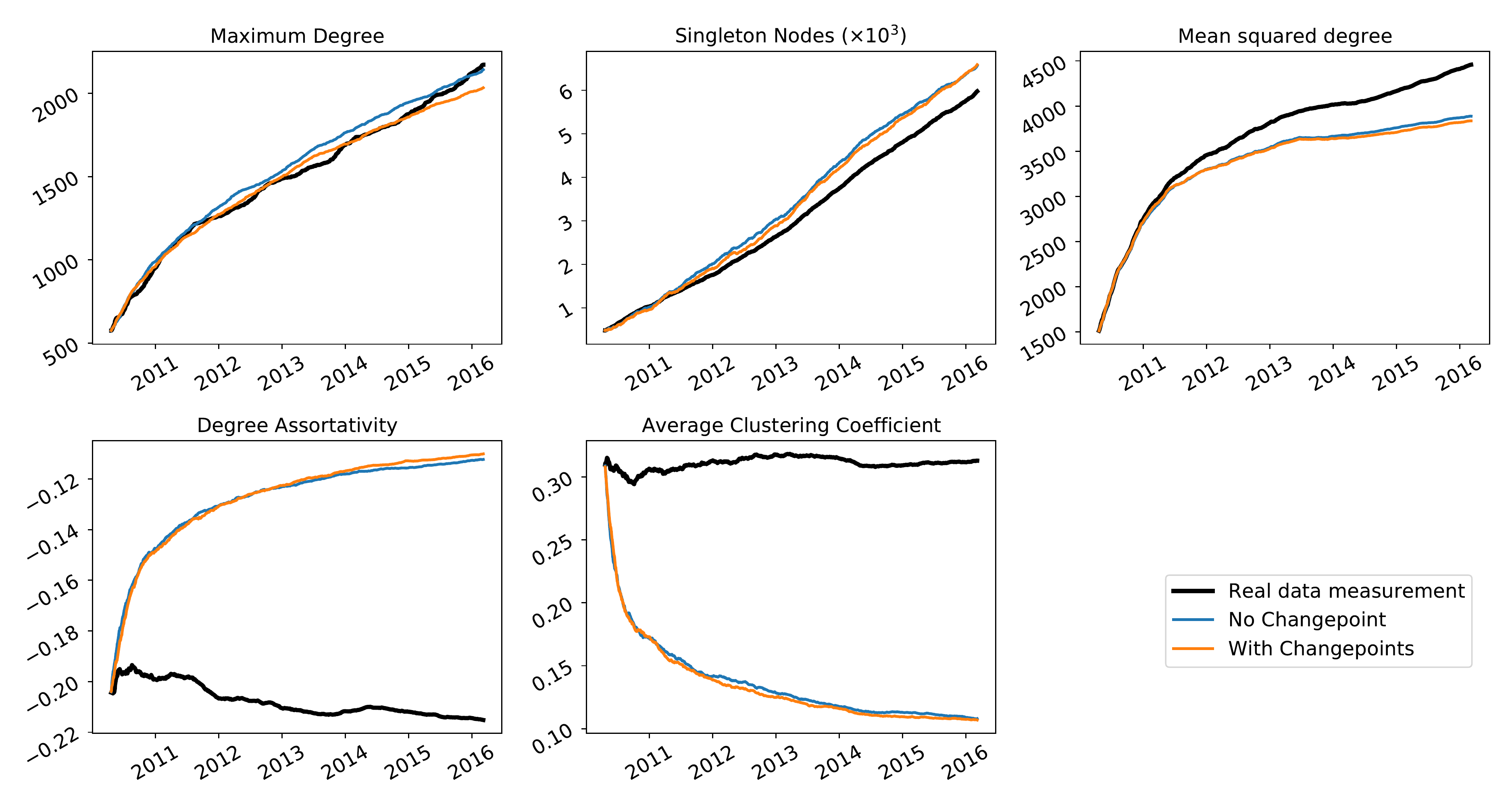}
	\caption{Stack Exchange MathOverflow: Comparison on various network statistics between best fitting mixture model without change and its best fitting counterpart fitted over 10 equally sized intervals for the Stack Exchange MathOverflow dataset.}
	\label{fig:SXcomparisons}
\end{figure*}

To demonstrate the relevance of this methodology, we investigate its use on four real world network datasets: the arXiv high energy phenomenology (cit-hepPh) citation network~\cite{gehrke2003overview}, a Facebook wall posts dataset~\cite{viswanath2009evolution}, a StackExchange MathOverflow dataset~\cite{paranjape2017} and the Enron email corpus~\cite{klimt2004enron}.
More information on the datasets is given in supplementary material.
Importantly, for the first three datasets, there isn't any prior reason for which we would expect to see behaviour that might be described as a changepoint, as there are no documented events to our knowledge that would change how nodes would interact in those networks.
The Enron dataset, however -- the corpus of emails between company employees made public when it was being investigated -- spans the 2001-2002 period in which many exogeneous events occurred that one might expect to influence connection forming within the network.
For each dataset, we fit an object model of the form
\begin{equation}\label{eqn:realdatafits}
M(t) = \beta_{\text{BA},F(t)} M_{\text{BA}} + \beta_{\text{tri},F(t)}M_{\text{tri}} + \beta_{\text{rand},F(t)} M_{\text{rand}}
\end{equation}
to the dataset, where, as with the synthetic network example, we use $J$ evenly sized time intervals and $F(t)$ corresponds to the interval number containing time $t$.
These three models represent three different processes that may contribute to network growth: a tendency to connect to nodes of higher degree, a tendency to connect node pairs that have mutual connections and a random factor.
To find a good number of time intervals to use, we see how the likelihood varies going from just a single time intervals up to 18 evenly spaced time intervals, using the $c_0$ value defined in the Likelihood calculation section.
A model of the form~\ref{eqn:realdatafits} is fitted for each number of time intervals and the $c_0$ value calculated for the best fitting model in each case, and these are shown in Figure~\ref{fig:c0vscps}, normalised by its maximum value in the right hand plot.
We see from the right hand plot on Figure~\ref{fig:c0vscps} that the model for the Enron dataset benefits the most from the addition of changepoints, with the increasing $c_0$ trend seen throughout the whole range of number of changepoints used, compared to the other datasets whose $c_0$ flattens off very quickly.
The fits found for each of these are displayed in Figure~\ref{fig:enrontime}.
From Figure~\ref{fig:enrontime} we can see that all four networks have a reasonably high BA component as well as a triangle-closure based component, with the random model seeming to fill the gaps.
In the case of arXiv and StackExchange the components keep very similar proportions across time.
In the case of Facebook we can see a large increase in the random component in the later data.
In the case of the Enron data, we have added the times of some of the documented events of the scandal for context, though we do not assert that movements in the data are caused by these events, especially when just using ten time intervals.
This being said, we see that there is a small peak in the degree (BA) model component proportions around the time of the bankruptcy announcement (event 2) when the central figures of the network are contacted by many different individuals in the few days following this.
Finally, we generate artificial networks using these model fits and the operation model extracted from the data to investigate if there is an improved reproduction of network statistics by allowing a time-varying component.
Figures~\ref{fig:ENcomparisons} and~\ref{fig:SXcomparisons} show various network statistics for the real network and the networks generated by the best fitting model, for a single time interval $J=1$ and $J=10$ time intervals, denoted ``no changepoint" and ``with changepoints" respectively (results just for the StackExchange and Enron networks are shown, those for the two remaining datasets are in the supplementary information).
Results from the models show the average value from 10 realisations, and 95\% confidence intervals.

One thing that can be concluded from these real data results is that the fitted models do not well reproduce the majority of network statistics considered for these real networks. Given the fitted models were the highest likelihood combination of the three input models this gives us reasonable certainty that no combination of these input models can provide a good fit to this set of statistics. This should not be a surprise since the real graphs are the results of highly complex processes between individuals. It should be considered a strength of our technique. There is no other technique (that the authors know of) that could rule out the idea that some combination of those models at different time intervals would well explain the data. The only alternative we know would be growing and measuring a network for each combination of parameters at each time interval and this would be infeasible computationally.

Given that Figures~\ref{fig:ENcomparisons} and~\ref{fig:SXcomparisons} suggest only modest improvements to network statistics reproduction by including changepoints, we may ask whether the inclusion of more time intervals leads to a significant increase in the model likelihoods.
We check this using Wilks' theorem~\cite{wilks1938large} on the FETA model as explained in~\cite[Example 3.1]{clegg2016likelihood}.
We are fitting an object model of $L$ different components over $J$ time intervals, so let $\beta_{lj}$ be the proportion of the $l^{\text{th}}$ model component at time interval $j$, with $\sum_{l=1}^L \beta_{lj} = 1$ for all $j$.
We then test $H_0 \colon \beta_{ij} = \beta_i$, i.e. a model with no changepoints, against $H_1 \colon \beta_{lj}$ varies with $j$. Wilks' theorem allows us to test the statistical significance of $H_1$ vs $H_0$ (accounting for the number of extra parameter in $H_1$). The same procedure can be used to compare adding more time intervals if $H_0$ ``nests" inside $H_1$, that is $H_1$ has the same changepoints as $H_0$ and some extra changepoints (alternatively this can be thought of as $H_1$ contains all the time intervals in $H_0$ and subdivides one or more). Running this test on the four datasets in this paper shows that adding one or more changepoints where there were none before is statistically significant at $p < 0.0001$. Again in all four datasets moving from two to ten timeintervals was statistically significant with $p < 0.0001$. In all data sets but StackExchange, moving from five to ten time intervals was statistically significant with $p < 0.01$ (this dataset was the one that showed the smallest change in $\beta_{lj}$ as $j$ varies, see figure~\ref{fig:SXcomparisons}).


In the StackExchange dataset, even without adding changepoints, the model mixture captures most statistics fairly well apart from the clustering coefficient, and adding changepoints makes a modest improvement to most of these statistics.
The latter is true for the Enron emails dataset, though the difference is very small.
It is clear that these relatively simple models are not correctly reproducing all network statistics (that is not the aim of this paper).
However, it can be seen that adding changepoints has increased the fidelity of the model for reproducing most of the statistics studied.
Further work will investigate adding other model components to the mixture and how well this improves the ability of artificial models to recreate network statistics.

\section*{Discussion and conclusion}
This work shows a way forward for the promising research area of growth network models that change in time. The framework given here allows a large number of existing models to be combined with time-varying parameters. This can generate a large family of different networks according to the user's need. When investigating real data we can fit models that change in time and look for points in the network history when the underlying model generating the network data changes. We showed how this can reveal how and when the importance of different growth mechanisms changes over time in real networks. 

This adds to a toolkit of techniques for understanding time-varying networks, in a time where timestamped network data is becoming more available. Furthermore, it formally generalises and combines ideas from a wealth of literature on network growth models into a framework that is amenable to statistical inference. The main modelling focus in this work was on which nodes attract new links in growing networks, in isolation from other considerations such as the frequency of new link/node arrivals. Our approach could be extended to addressing this, investigating the interplay between these aspects of network evolution. Whilst for simplicitly we treated all networks as undirected, the method works also for directed networks by making some minor modifications to object models, such as models dependent on node degree being instead dependent on in, out or total degree.

Using experiments on artificial data we have shown that we can not only capture the known underlying model that generated the observed network data but also that we can capture with good accuracy the point in time at which that model changed. We use a similarity measure to formally capture the notion of two growth models that may give rise to similar networks and show that the problem of estimating network parameters or change points is more difficult when they are more similar. The good performance of our estimation procedure gives us confidence that the framework can accurately evaluate different explanatory models for real data and how those models may change throughout the observed period in a real dataset. 

We investigated four well-known data sets from the literature and use three common mechanisms applied in network models. We show that for some datasets the proportion of these models can change considerably over time while in other data sets the proportions remain largely fixed. The mechanisms used here are deliberately chosen to be simple and therefore do not accurately generate all network statistics. However, we show that in some cases adding models that vary in time can improve this. There is a potentially large field of research in investigating model components that could be used to better capture these statistics. 

The code we used in this paper is available on GitHub as multipurpose software\footnote{https://github.com/narnolddd/FETA3} for generating networks from a given object and operation model and fitting mixed and time varying models to real data.

\bibliographystyle{unsrt}
\bibliography{SRRevised/biblio}

\end{document}


\maketitle

\section{Likelihood Calculation}\label{sec:likecalc}

The main text describes the general form of a model likelihood given observations of an evolving graph $G_1, G_2, \dots, G_t$:
\begin{equation}\label{eqn:likelihoodprod}
l(M| G=g) = \prod_{i=1}^t \mathbb{P}(\varDelta_i = \delta_i | G_{i-1}=g_{i-1}, M).
\end{equation}
Here we provide more details for how each term $\mathbb{P}(\varDelta_i = \delta_i | G_{i-1}=g_{i-1}, M)$ is evaluated. In all of the data sets used in this paper, the graphs evolve by adding stars. For example, in the arXiv citation dataset the graph evolves when a new paper arrives and it cites other papers in the dataset. The operation model here is a new node arriving and connecting to $m$ existing nodes. In this case then the $m$ nodes of the star are chosen in turn, all without replacement, to avoid self-loops or multilinks. Now, to calculate the exact probability of this operation we must evaluate the probability of each order occurring. So if we want the likelihood that the chosen set of nodes was $\{2,3\}$ then we must consider the likelihood of picking 2 then 3 and also 3 then 2. 

\begin{figure}[htbp]
    \centering
    \includegraphics[width=0.4\linewidth]{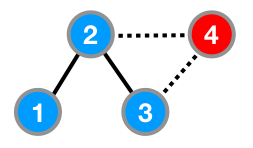}
    \caption{A new node (4) joining the network and connecting to two existing nodes (2,3).}
    \label{fig:newnodelikelihood}
\end{figure}

Consider the observation in figure~\ref{fig:newnodelikelihood}, where we observe a new node (4) joining the network by connecting to nodes (2) and (3), and suppose we wish to compare which of two object models, $M_1$ and $M_2$, is a better explanation for this observation.
As (4) is a new node, the probability we are interested in is the probability of picking nodes (2) and (3) according to models $M_1$ or $M_2$. As there is no order on which of the edges $(4,2)$ and $(4,3)$ arrived first, we consider the two different orders. Let us imagine we want compare the BA model with the random model for the pictured addition of one node and two links. In general we have
\begin{equation*}
    l(M|\text{observation}) = \mathbb{P}(\text{pick node 2, then 3}|M) + \mathbb{P}(\text{pick node 3, then 2}|M).
\end{equation*}
For the BA model we get the following likelihood
\begin{align*}
    l(M_{\text{BA}}|\text{observation}) &= \underbrace{\frac{k_2}{k_1 + k_2 + k_3} \cdot \frac{k_3}{k_1 + k_3}}_{\text{node 2 then node 3}}+ \underbrace{\frac{k_3}{k_1 + k_2 + k_3} \cdot \frac{k_2}{k_1 + k_2}}_{\text{node 3 then node 2}} \\
    &= \frac{1}{2} \cdot \frac{1}{2} + \frac{1}{4} \cdot \frac{2}{3} \\
    &= \frac{5}{12}
\end{align*}
and for the random model we get 
\begin{align*}
    l(M_{\text{rand}}|\text{observation}) &= \frac{1}{3} \cdot \frac{1}{2} + \frac{1}{3} \cdot \frac{1}{2} \\
    &= \frac{1}{3}.
\end{align*}
In this case, therefore, the BA model is more likely (the likelihood ratio of BA to random being $\frac{5}{4}$. If a node connects to a small number of others then each ordering can be considered explicitly in this way. However the number of orderings goes up quickly with the number of chosen nodes $m$ (the number of possible orderings of node selections being $m!$) and hence, for $m > 5$ we use a sampling procedure to calculate the average likelihood for a sample of possible orderings and multiply this by the number of orderings. 
\section{Data requirements for the framework}\label{sec:datreq}
Our framework requires that the history of the network's growth is known to a sufficient temporal resolution such that the edges arriving in each time increment form a star graph. The simplest scenario is if just a single link arrives at each time increment; this is the case for the Facebook wall posts and StackExchange MathOverflow datasets. In many datasets, including all our artificial datasets, each time increment will comprise link arrivals as a star (an existing or new node connecting to more than one other new or existing nodes simultaneously). This is subject to the ordering problem discussed in the previous section, in principle the calculation can be exact but the combinatorics mean that sampling needs to be used. In some data sets more complex situations arise, for example, two unrelated links arriving simultaneously. In principle these could be handled by considering every possible arrival order as well. However, in the data sets we have used these events are extremely rare. They are less than 0.1\% of the data and the ordering makes negligible difference to the likelihood calculated so an arbitrary ordering can be assumed without change to the results. Data sets where a large number of unrelated links arrive at exactly the same instant cannot be analysed with this methodology but this still leave a large and increasing number of datasets amenable to analysis. 
\section{Datasets}
This paper used four different publicly available network datasets as case studies for fitting (time varying) mixtures of models. To make these compatible with the modelling framework, we cleaned the datasets to remove any duplicate links and any nodes which do not connect to the largest connected component upon joining the network. These cleaned datasets are available within the software repository~\cite{Arnold2020} in the form of a tab separated file, where each line is of the form \begin{verbatim}
SOURCE_NODE    DEST_NODE   TIMESTAMP
\end{verbatim}
specifying a string source and destination node for each link and the timestamp (Unix epoch) at which the link was recorded. Table~\ref{tab:datasets} details each dataset showing the original source and number of nodes/edges after cleaning. Also included is a description of what an edge $(u, v, t)$ between nodes $u$ and $v$ created at time $t$ represents for this network. For simplicity we consider these networks as undirected.
\begin{table}[htbp]
    \centering
    \begin{tabular}{p{3cm} p{9cm} c c}
        \hline
        \textbf{Name} & \textbf{Description and source} & \textbf{Nodes} & \textbf{Edges} \\
        \hline
        Enron emails & Network constructed by a dataset of emails sent or received by Enron employees, where a node is an individual who has emailed or been emailed by an Enron employee. Edge $(u,v,t)$ represents $u$ emailing $v$ at time $t$. Hosted at \url{http://konect.cc/networks/enron/}~\cite{klimt2004enron} & 86978 & 297456 \\
        \hline
        ArXiV Cit-Hep-Ph & Citation network of papers uploaded to ArXiV under the Physics High-Energy Phenomenology tag between 1993 and 2003. Edge $(u,v,t)$ represents paper $u$ being uploaded at time $t$ citing paper $v$. Hosted at \url{https://snap.stanford.edu/data/cit-HepPh.html}~\cite{gehrke2003overview} & 34343 & 412365 \\
        \hline
        Facebook wall posts & Interaction dataset from wall posts between Facebook users in the New Orleans area over roughly 2.5 years (2006-2009). Edge $(u,v,t)$ represents user $u$ posting on $v$'s wall at time $t$. \url{http://konect.uni-koblenz.de/test/networks/facebook-wosn-wall}~\cite{viswanath2009evolution} & 45813 & 183412 \\
        \hline
        StackExchange MathOverflow & Dataset comprising interactions on the MathOverflow forum up to March 2016. An edge $(u,v,t)$ exists if at time $t$ user $u$ answers $v$'s question, replies to $v$'s question or comments on $v$'s answer to a question. \url{https://snap.stanford.edu/data/sx-mathoverflow.html}~\cite{paranjape2017} & 24759 & 187985 \\
        \hline
    \end{tabular}
    \caption{Description of datasets used in the paper.}
    \label{tab:datasets}
\end{table}
\section{Extra dataset results}
\begin{figure}[htbp]
    \centering
    \includegraphics[width=0.9\linewidth]{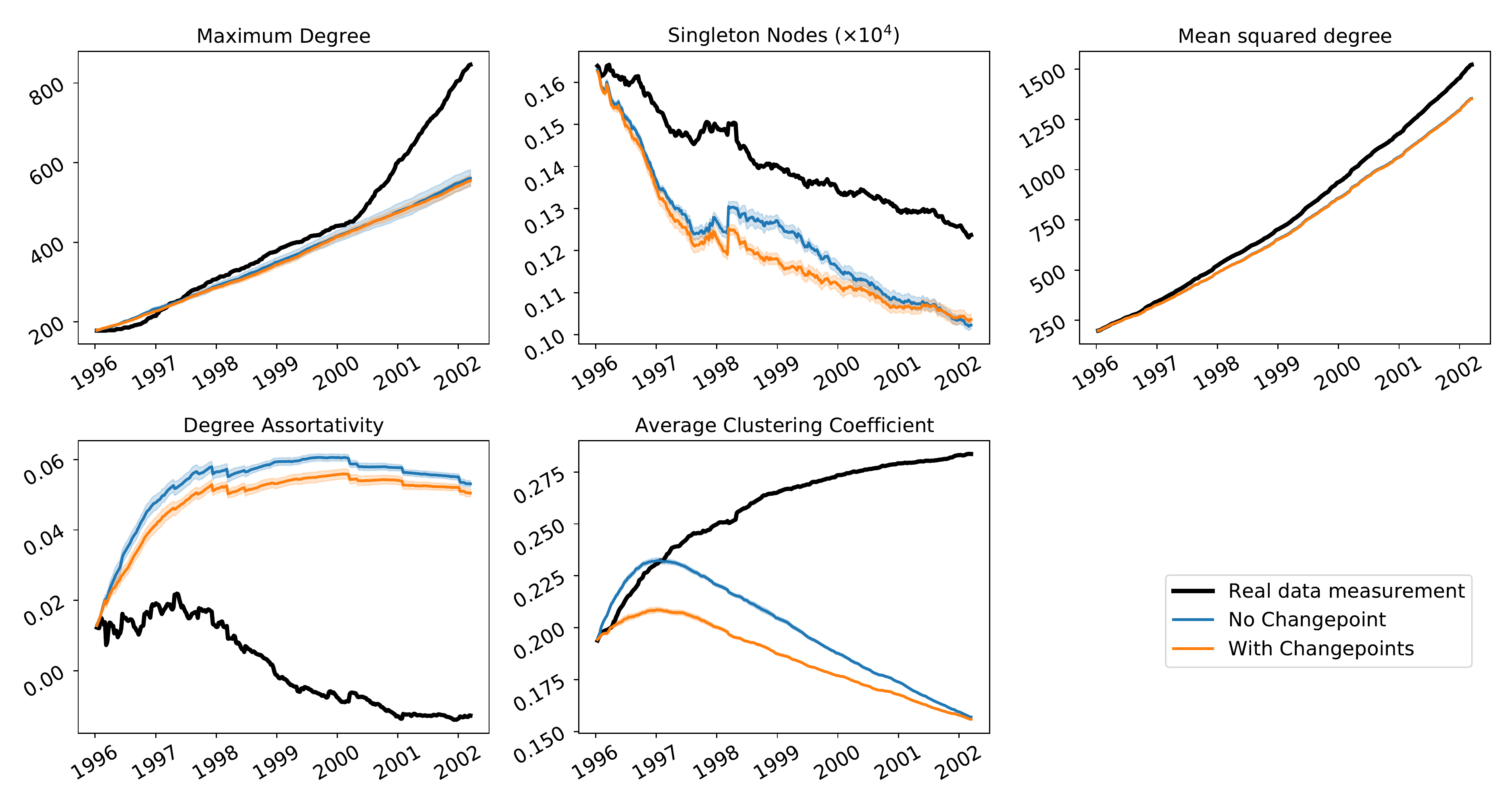}
    \caption{ArXiV Cit-Hep-Ph: Comparison on various network statistics between best fitting mixture model without change and its best fitting counterpart fitted over 10 equally sized intervals for the citation network dataset. }
    \label{fig:cit}
\end{figure}

\begin{figure}[htbp]
    \centering
    \includegraphics[width=0.9\linewidth]{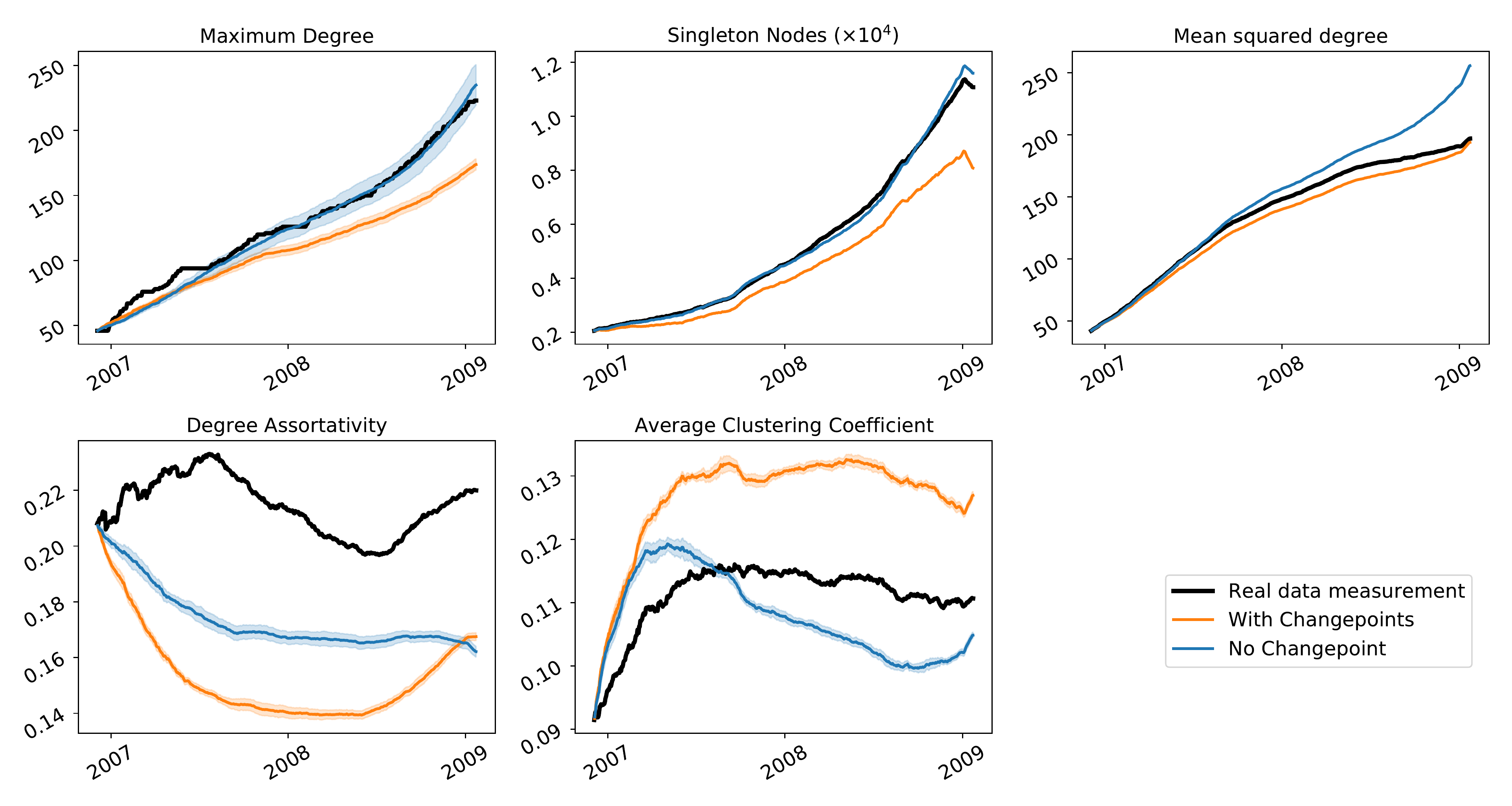}
    \caption{Facebook wall posts: Comparison on various network statistics between best fitting mixture model without change and its best fitting counterpart fitted over 10 equally sized time intervals for the Facebook wall posts dataset. }
    \label{fig:fb}
\end{figure}

As well as the real data results on figures 8 and 9 with the experiments described in the main text, we generated networks for the remaining Facebook wall posts and citation network datasets. In the case of the citation network (Figure~\ref{fig:cit}), the `first order' degree based statistics (top row) from the artificial networks are fairly close to the real data but adding changepoints doesn't seem to add any advantage. The last point is not surprising; there was not much change in the model parameters in the time-varying model (Figure 9 main text). The degree assortativity and clustering coefficient are not realised well by either model. This being said, the model components used for this were deliberately simple to be able to compare the four networks as in Figure 9 main text; if the aim was to find the most realistic model, more complex models such as nonlinear preferential attachment~\cite{krapivsky2000connectivity} or aging~\cite{dorogovtsev2000evolution} could be considered. For the Facebook wall posts network (Figure~\ref{fig:fb}), the degree-based statistics were well reproduced by both models, with the changepoints not adding much advantage, while the degree assortativity and clustering coefficient were poorly captured by both models. We suggest here that a more realistic model may be achieved by incorporating community structure such as a dynamic variant of the stochastic block model.
\section{FETA code for reproducing the experiment in the manuscript}
We have provided a Java-based codebase FETA~\cite{Arnold2020} (Framework for Evolving Topology Analysis) for implementing various aspects of this framework, with tutorials provided. The user can generate a network given an operation and object model, calculate the likelihood of a model given network observations (i.e. a dataset of fitting the requirements in section~\ref{sec:datreq}), fit a time-varying mixture model to network observations, and extract a time series of different network measurements from a dataset. An API is also provided for users to write their own object models to test on data.

\bibliographystyle{unsrt}
\bibliography{SRRevised/biblio}